\documentclass{WileyMSP-template}
\usepackage[
square,super,numbers,round, comma, sort&compress]{natbib} 
\usepackage{rotating}

\usepackage{amsmath,amssymb}

\usepackage{array,makecell,multirow}
\usepackage{placeins}
\usepackage{graphicx} 
\usepackage[justification=justified]{caption}
\usepackage{wrapfig, framed, caption}
\usepackage[export]{adjustbox}
\usepackage{stfloats}
\usepackage{caption}

\usepackage{geometry}
\usepackage{amsmath}
\usepackage{siunitx}

\begin{document}

\pagestyle{fancy}
\rhead{\includegraphics[width=2.5cm]{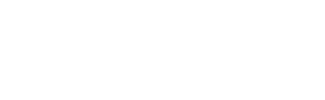}}

\title{
Nanocolumnar Material Platforms:Universal structural parameters revealed from optical anisotropy }

\maketitle


\author{Ufuk Kilic*}
\author{Yousra Traouli}
\author{Matthew Hilfiker}
\author{Khalil Bryant}
\author{Stefan Schoeche}
\author{Rene Feder}
\author{Christos Argyropoulos}
\author{Eva Schubert}
\author{Mathias Schubert*}



\begin{affiliations}
U. Kilic, Y. Traouli, K. Bryant, C. Argyropoulos, E. Schubert, M. Schubert\\
Department of Electrical and Computer Engineering, University of Nebraska-Lincoln, Lincoln, NE 68588, USA\\
E-mail address: ufukkilic@unl.edu\\
M. Hilfiker\\
Onto Innovation Inc., Wilmington, MA 01887, USA\\
S. Schoeche\\
J.A. Woollam Co., Inc., Lincoln, NE 68508, USA\\
R. Feder\\
Fraunhofer Institute for Microstructure of Materials and Systems (IMWS), 106120, Halle (Saale), Germany\\
C. Argyropoulos\\
Department of Electrical Engineering, The Pennsylvania State University, University Park, PA 16803, USA\\
M. Schubert\\
Solid State Physics and NanoLund, Lund University, P.O. Box 118, 22100, Lund, Sweden
\end{affiliations}




\begin{abstract}
\fontsize{12pt}{14pt}\selectfont


Nanostructures represent a frontier where meticulous attention to the control and assessment of structural dimensions becomes a linchpin for their seamless integration into diverse technological applications. 
By using integrative and comprehensive methodical series of studies, we investigate the evolution of the depolarization factors in the anisotropic Bruggeman effective medium approximation, that are extremely sensitive to the changes in critical dimensions of the nanostructure platforms. 
To this end, we fabricate spatially coherent highly-ordered slanted nanocolumns from zirconia, silicon, titanium, and permalloy on silicon substrates with varying column lengths using glancing angle deposition. 
In tandem, broad-spectral range Mueller matrix spectroscopic ellipsometry data, spanning from the near-infrared to the vacuum ultraviolet (0.72~eV to 6.5~eV), is analyzed with a best-match model approach based on the anisotropic Bruggeman effective medium theory. We thereby extracted the anisotropic optical properties including complex dielectric function, birefringence, and dichroism. Most notably, our research unveils a universal, material-independent inverse relationship between depolarization factors and column length. 
We envision that the presented universal relationship will permit accurate prediction of optical properties of nanocolumnar thin films improving their integration and optimization for optoelectronic and photonic device applications.


\end{abstract}


\section{Introduction}
Within the last two decades, light-matter interactions at nanoscale induced by artificial nanostructures have attracted great interest in numerous areas of physics, medicine, biology, chemistry, and pharmacology.\cite{krishnamoorthy2012topological,segal2015controlling,liu2015highly,mazaleyrat1981chiral,jahani2016all, wang2017circular, gansel2009gold,  kuzyk2012dna, zhang2008superlenses, cao2012valley, rodrigues2014nonlinear, zhang2014electrically, kang2015active} As a promising nanofabrication technique, glancing angle deposition (GLAD) offers the creation of spatially coherent, highly porous, super-lattice type three dimensional nanomorphologies.\cite{kasputis2013slanted,rice2017control, kaschke2016optical,kasputis2015use, schmidt2013generalized,ye2005uniform,gibbs2013plasmonic,mark2013hybrid,tasco2016three,gansel2012tapered, sarkar2019shape,singh2013wafer,robbie1999chiral,kilicc2019tunable, kilic2021broadband,sekora2017optical} The use of particle flux at oblique angle of incidence in combination with a precise sample manipulator in this bottom-up  physical vapor deposition technique enables the fabrication of versatile 3-D nanostructures over wafer scale substrates. In comparison with other top-down approaches such as lithography techniques, GLAD is feasible, time, and cost effective.\cite{ariga2008challenges, kilicc2019tunable, kilic2021broadband} Moreover, since this technique can control structural parameters including geometry and porosity, it is also possible to tailor and fine-tune the optical and material properties of nanostructures.\cite{schmidt2013anisotropic} Due to the complexity of optical anisotropy possessed by these highly porous nanostructures, performing the growth-time dependent critical dimension analysis and unraveling the correlation between the optical and structural parameters remain challenging. A robust method for accurately determining aforementioned information is a necessary component in reliable evaluation and optimization of the nanostructure device performance, for example.

\begin{figure}[htb]
\centering
\includegraphics[width=0.65\linewidth]{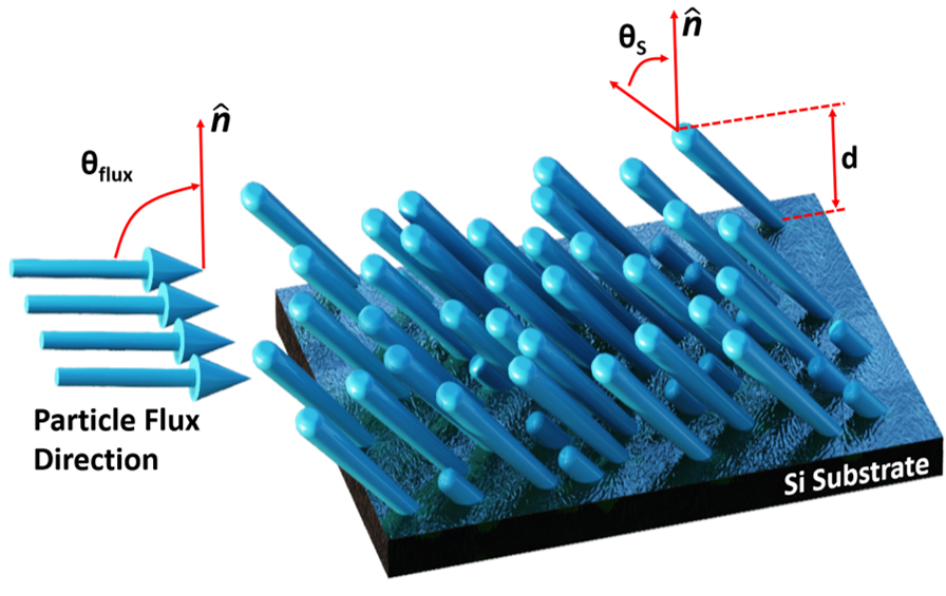}%
    \caption{(a) Schematic illustration of columnar growth process using the GLAD technique. $d$ is the thickness of slanted columnar structure. The depicted angles: $\theta_{flux}$ is the angle between the surface normal ($\hat{n}$) and the particle flux direction. $\theta_{s}$ is the slanting angle of the columns relative to the surface normal. }
\label{Fig1_Growth}
\end{figure}

Mueller matrix a.k.a. generalized spectroscopic ellipsometry (SE) has been demonstrated as an effective, non-destructive, and model-based characterization method to investigate the intrinsic structural parameters and anisotropic optical properties of slanted columnar thin films (SCTFs) (see the schematic illustration of an SCTF in Figure~\ref{Fig1_Growth}).\cite{schmidt2009optical,gospodyn2006characterization,liang2013generalized,nerbo2010characterization} In order to precisely identify the structural parameters and effective optical constants, appropriate physical models are required to analyze the generalized SE experimental data measured on SCTFs. For example, Nerb{\o} \textit{et al.} performed the characterization of inclined GaSb nanopillars by using Mueller matrix SE approach and applied a simplistic structural graded anisotropic effective medium model.\cite{nerbo2010characterization} Schmidt~\textit{et al.} reported the performance of different anisotropic Bruggeman effective medium approximation (AB-EMA) based homogenization approaches for columnar thin films and compared results with those obtained with the homogeneous biaxial layer approach (HBLA).\cite{schmidt2009optical} The use of the HBLA method in the model analysis of generalized SE data assumes the optical response of SCTFs as a single homogeneous layer. This assumption allows the wavelength-by-wavelength determination of effective dielectric functions along three major Cartesian dielectric polarizability axes, ($\mathrm{a}$,~$\mathrm{b}$,~$\mathrm{c}$; Figure~\ref{Fig2_schematics_AB-EMA}) which thereby form an orthorhombic system. While the HBLA model also provides the wavelength-independent structural parameters: slanting angle of columns ($\theta_{\mathrm{s}}$) and thin film thickness ($d$), the AB-EMA model provides the depolarization factors representing the structural dimensions along each of the major polarizability axes. Furthermore, the AB-EMA provides an effective isotropic bulk dielectric function that represents the isotropic bulk material from which the nanocolumns are constructed.\cite{Schmidt_2013,hofmann2017screening} 

In another report, Schmidt \textit{et al.} presented the optical properties of cobalt SCTFs passivated by atomic layer deposition of Al$_{2}$O$_{3}$ ultra-thin film which were successfully extracted by using AB-EMA based generalized SE data analysis.\cite{schmidt2012optical} Liang  \textit{et al.} applied AB-EMA to Mueller matrix SE data of porous slanted columnar thin films infiltrated with polymer (poly(-methyl methacrylate)). In their study, in addition to the spectral evolution of dielectric function, the changes in both void fraction rate and the slanting angle together with the thin film thickness were reported.\cite{liang2013generalized}  Similar to this study, Kasputis \textit{et al.} monitored \textit{in-situ} the polyacrylic acid polymer brushes grafted onto SCTFs using a combined quartz crystal microbalance and generalized SE technique. Kilic~\textit{et al.} and Sekora~\textit{et al.} also utilized the HBLA in the extraction of optical constants of Si-Au  and Co-NiFe  slanted columnar heterostructure thin films.\cite{kilic2018tailoring,sekora2017optical} Hofmann~\textit{et al.} investigated the Terahertz response of metal SCTFs and derived resistivity and free charge carrier properties.\cite{hofmann2017screening} Previously, the depolarization parameters were reported without much discussion. Specifically, a detailed study of the effect of the columnar geometry onto the depolarization factors has not been performed. A detailed discussion of the effect of fundamental material properties such as those differentiating dielectrics from semiconductors, or semiconductors from metals, have not been comparatively discussed. It is rather interesting whether or not the AB-EMA provides a material-independent method to cast the resulting anisotropic dielectric function of a given SCTF, regardless of the condensed matter properties of the material constituting the columnar structure. In this study, we report the evolution of the depolarization factors as a function of SCTF growth time, and thereby as a function of column length, or SCTF thickness. We perform a series of systematic thickness variations in GLAD deposited zirconia, silicon, titanium, and permalloy SCTFs deposited on Si substrates. We use Mueller matrix SE over a large spectral range from the near infrared ($\approx$0.72~eV) to the vacuum ultraviolet ($\approx$~6.5~eV). We use the AB-EMA formalism and report the evolution of the depolarization factors as a function of columnar length for all materials investigated. We suggest a simple asymptotic relationship for the dependencies of the depolarization factors on the thickness of SCTFs. We find these relationships to be universally valid for all materials studied where the asymptotic model parameters may vary across different materials. We discuss the asymptotic behavior in view of the one-dimensional character of the columnar structures. This key output is valid regardless of both spectral range and the underlying physical mechanisms which produce the respective dielectric function behaviors within the near-infrared to vacuum ultra violet spectral regions in the dielectric, semiconductor, and metal nanostructures.





\section{Results and Discussion}

By applying the AB-EMA model for the analysis of measured Mueller matrix SE data, the anisotropic dielectric function spectra for ZrO$_{2}$, Si, Ti, and NiFe SCTFs are obtained, and presented in Figure~\ref{Fig_dielectric}~(a), (b), (c), and (d), respectively. The corresponding constituent effective \textit{bulk-like} dielectric functions obtained during the same model analysis are shown in Figure~\ref{Fig_bulkdielectric}. In general, the \textit{bulk-like} isotropic dielectric functions have similar physical line shapes when compared with the effective anisotropic dielectric function for polarization along direction $\mathrm{c}$ (solid lines in Figure~\ref{Fig_dielectric}). Similar observations, albeit over much smaller spectral regions were made previously, e.g., for Co, Ti, and Si SCTFs by Schmidt and Schubert.\cite{schmidt2013anisotropic} 

Importantly, such columnar nanostructures exhibit extreme birefringence and dichroism properties.\cite{zhang2022high,kilic2018tailoring,schmidt2013anisotropic,gu2015color,ray2023birefringent} 
These intrinsic optical properties play fundamental roles in modulating and controlling the light-matter interactions. While the polarization selective attenuation (in some contexts, extinction) is mainly related to the dichroism, the changes in retardation of incident beam along its different propagation directions in the media is mainly related to the birefringence property. Hence, the ability to tailor these optical properties is also important for their potential use in next generation waveguide and sensing technologies, biological imaging, spintronics, nano-scale electronics, high efficient and tunable filters, and beam splitters, for example.\cite{noyan2022birefringence, zhang2022cavity,kovalev2001strong, basiri2019nature,solomon2020nanophotonic,ferreira2021nanostructured,mourik2012signatures} However, to the best of our knowledge, a systematic experimental investigation of both birefringence and dichroism in glancing angle deposited nanostructures is rarely reported \cite{liang2013generalized} and the dependency of these properties on nanocolumnar thickness and wide variety of material types have not yet been reported. With the extraction of anisotropic dielectric functions via AB-EMA based best match model calculations of Mueller matrix SE data (see Figure \ref{Fig_dielectric}), it is possible to reveal these optical anisotropies, as well. 
Hence, these optical properties were extracted in terms of anisotropic refractive index and extinction coefficient based on the following formula; 
\begin{equation}
\Delta(n,k)_{eff}=((n,k)_{\mathrm{c}}-((n,k)_{\mathrm{a}}+(n,k)_{\mathrm{b}})/2))
\label{eq_BireDich}
\end{equation}
where $\Delta~n_{eff}$ and $\Delta~k_{eff}$ are the effective refractive index and extinction coefficient differences that represent the birefringence and dichroism, respectively.\cite{liang2013generalized} Figure \ref{Fig_Dneff_Dkeff} shows behavior of both birefringence and dichroism values at a selected photon energy values where the dielectric function along major polarizability axis (i.e. $\varepsilon_{c}$) takes its maximum value within the spectral range of interest (0.72~eV to 6.5~eV). Interestingly, both $\Delta~n_{eff}$ and $\Delta~k_{eff}$ values asymptotically saturate as the thickness increases. 
To provide a better quantitative assessment of these parameters, we also propose the following semi-empirical asymptotic relation for both of these parameters;
\begin{equation}
\Delta(j)_{eff}=\xi{j}-\frac{\xi_{\mathrm{j,0}}}{1+\eta_{j} d },
\label{Eq:bire_dichr}
\end{equation}

\noindent where j is either n or k. Equation \ref{Eq:bire_dichr} introduces an infinity-thickness asymptotic parameter $\xi_{j}$ (the value which the effective refractive index and extinction coefficient differences will approach at infinitely thick SCTF), the amplitude ($\xi_{j,0}$ ), and inverse slope parameter ($\eta_{j}$). The value of each parameters employed in Equation \ref{Eq:bire_dichr} is listed in the Table S1 in the supplementary material. As it is seen from Figure~\ref{Fig_Dneff_Dkeff}, for each material choices, the differentiability of the complex dielectric function along major polarizability axis ($\varepsilon_{\mathrm{c}}$) from the dielectric functions along the other two polarizability directions ($\varepsilon_{\mathrm{a}}$ and $\varepsilon_{\mathrm{b}}$) increases with thin film thickness and this climbing trend asymptotically slows down for thicker thin films. Interestingly, the proposed equation can resolve this behavior for any material choices investigated as a part of this study. 
The presence of different mechanisms (e.g. ad-atom diffusion, ballistic shadowing, formation of nucleation centers over the surface, mobility of atoms) in GLAD process can result in the spread-out and aberrations (see Figure \ref{Fig_Dneff_Dkeff}, blue circle symbols: experimental data points) from the main trend (see Figure \ref{Fig_Dneff_Dkeff}, red solid lines: generated based on semi-empirical fit function). Thereby, revealing the optical anisotropy information including both birefringence and dichroism of columnar nanostructures from various material choices lays a foundation for further study of optoelectronic applications and also provides the methodology for the investigation of these optical anisotropy types of other materials.

In addition to the anisotropic dielectric function, $\Delta n_{eff}$ and $\Delta k_{eff}$, AB-EMA framework provides another set of parameters, the depolarization factors ($q_{\mathrm{a}}$, $q_{\mathrm{b}}$, and $q_{\mathrm{c}}$), which are extremely sensitive to the size and shape of the material under investigation. Despite its critical importance, there is very seldom discussion over the effect of a systematic comparison of size changes on the evolution of these factors. However, the disparity of values depending on the various material types and thickness advances this work into a case-by-case analysis, where their predictive rules will also be provided. The extraction of such information is essential because the critical dimensions (e.g. radius, length, periodicity etc.) of especially photonic metamaterial platforms play fundamental roles in engineering their inherent material properties.\cite{mourik2012signatures,limonov2017fano,saha2022tailoring,liu2022high,ringe2012plasmon} Figure~\ref{Fig_depolarization} summarizes the thin film thickness dependent evolution of parameters $q_{\mathrm{c}}$ and $q_{\mathrm{ab}}$ for ZrO$_{2}$, Si, Ti, and NiFe SCTFs. We observe a mostly monotonic evolution of all parameters for all materials as a result of investigating numerous different SCTFs with different growth time resulting in different thickness of the individual samples. We observe that the ``$\mathrm{a}-\mathrm{b}$" split parameter is near approximate to but less than 0.5 while q$_{\mathrm{c}}$ is always less than at least halve of the split parameter. For all samples investigated in this work, the scenario $d.$ described in Sect.~\ref{sec:AB-EMA} is valid throughout here. Hence, the SCTFs form rods with largest extension along $\mathrm{c}$ and with elliptical cross section where the larger elliptical extend is along direction $\mathrm{b}$, i.e., within the slanting plane. This behavior can indeed be observed within the cross- and top view-sections of the scanning electron microscopy images in Figure~\ref{Fig2_Growth_SEMs}.

The monotonic behavior of $q_{\mathrm{ab}}$ and $q_{\mathrm{c}}$ can be understood because in the limit of $D_{\mathrm{c}}\rightarrow \infty$, we anticipate that $q_{\mathrm{c}}\rightarrow 0$. Therefore, the asymptotic limit for $D_{\mathrm{c}}$ is zero. However, an asymptotic limit for the ratio of the ellipsoidal shape parameters $D_{\mathrm{a}}$ and $D_{\mathrm{b}}$ cannot be ranged in further except that it cannot be zero or unity since then the nanostructures would correspond to sheets not to long rods. A clear monotonic reduction approaching
near zero can be observed for all $q_{\mathrm{c}}$ dependencies, regardless of SCTF material. In order to provide a quantitative access and future numerical reproduction of our results, we suggest the following asymptotic functional relationship between the depolarization parameters, $q_j$ and the SCTF thickness, $d$;

\begin{equation}
q_{\mathrm{j}}=A_{j}-\frac{q_{\mathrm{j,0}}}{1+  S_{j} d}\quad,\quad\mbox j=(`\mathrm{ab}',`\mathrm{c}')
\label{Eq:qab-qc}
\end{equation}

\noindent where we introduce infinity-thickness asymptotic parameter $A_j$ (the value which the polarization parameter will approach at infinitely thick SCTF), amplitude ($q_{j,0}$) and inverse slope parameter ($S_{j}$). At zero thickness, we require that $0<A_{j}+q_{j,0}<1$, and at infinite thickness we require that $0<A_{j}<1$. Hence, we set these conditions accordingly when matching Equation~\ref{Eq:qab-qc} to the data points obtained from the AB-EMA analyses. The split between the depolarization factors along $\mathrm{a}$- ($q_{\mathrm{a}}$) and $\mathrm{b}$- ($q_{\mathrm{b}}$) axes, ab-split parameter (q$_{\mathrm{ab}}$), is calculated by using Equation \ref{eq:qab}. 
The proposed semi-empirical relations for both q$_{\mathrm{ab}}$ and q$_{\mathrm{c}}$ parameters can reveal the nonlinear common traces of each depolarization factors. While the aspect ratio (i.e., the ratio of columnar length to the column diameter) increases with the thin film thickness, the depolarization factor along \textit{c}-axis asymptotically approaches to its saturation value (i.e., A$_{q_{\mathrm{c}}}$, see Equation\ref{Eq:qab-qc}). Importantly, we used only $q_{\mathrm{ab}}$ and $q_{\mathrm{c}}$  parameters in the AB-EMA model analysis of Mueller matrix SE data. Especially for ZrO$_{2}$, Si, and Ti SCTFs, the ab-split value decreases with the increase in the thin film thickness. However, for the case of NiFe SCTFs, we found out a minor increase in the ab-split value. 
Due to the alloying nature of this magnetic material might also result in different diffusion dynamics and therefore might increase the thin film non-idealities (e.g., bifurcation, competition between columns and frequent column extinction) in the deposited columnar structures. 

Table~S-II in the supplementary material summarizes the best-match calculated parameters for the semi-empirical relations (Equation\ref{Eq:qab-qc}) for $q_{\mathrm{ab}}$ and $q_{\mathrm{c}}$ representing ZrO$_{2}$, Si, Ti, and NiFe SCTFs. As it is seen from  Figure~\ref{Fig_bulkdielectric}(d), the real part of dielectric function of NiFe SCTFs takes negative value in the substantial portion of the spectrum which is known as a clear signature of its metallic behavior. However, while the real part of dielectric function of Ti SCTFs is significantly smaller compared to the NiFe case, both ZrO$_{2}$ and Si SCTFs have positive dielectric functions (see Figure~\ref{Fig_bulkdielectric}(a)-(c)). Hence, for NiFe SCTFs, the screening effect result in one dimensional characteristics of the electric displacement vector that occurs at shorter column lengths. This leads the parameter $q_{\mathrm{c}}$ of NiFe SCTF to reach the saturation faster than the other material choices (see Figure~\ref{Fig_depolarization}). Whereas, slower growth-time dynamics of ZrO$_{2}$, Si, and Ti possibly cause the occurrence of asymptotic saturation in the depolarization factor along $\mathrm{c}$-axis for longer column length as compared with NiFe SCTFs. 
For example, for ZrO$_{2}$ SCTF, as the length of column is $\approx$250 nm, $q_{\mathrm{c}}$ value approaches to a constant value, but for NiFe SCTF case, the saturation occurs around $\approx$80 nm. The obtained common traces from Equation\ref{Eq:qab-qc}, $q_{\mathrm{c}}$ and $q_{\mathrm{ab}}$ (red solid lines in Figure~\ref{Fig_depolarization}) suggest that different saturation curves might stem from the material dependent differences in both ballistic shadowing and surface diffusion dynamics.

Using the cross section HR-SEM images of fabricated columnar nano-structures, the evolution of diameter along $\mathrm{b}$-axis as a function of the corresponding column length was obtained for each material and plotted in Figure~\ref{figure_qb} (red square symbols). In addition to this, Figure~\ref{figure_qb} also shows the column length dependent depolarization factor along $\mathrm{b}$-axis ($q_{\mathrm{b}}$) (blue square symbols) together with the retrieved curve for $q_{\mathrm{b}}$ (blue solid line) from its semi-empirical relation given in Equation\ref{Eq:qa-qb}. We can clearly observe the correlation between diameter along $\mathrm{b}$-axis and its relevant depolarization factor behavior. 
For each material choice, unlike the asymptotically decreasing behavior of parameter ${ q }_{ \mathrm{c} }$ (see Figure~\ref{Fig_depolarization}), parameter ${ q }_{ \mathrm{b}}$ (see Figure~\ref{figure_qb}) is nearly asymptotically increasing proportional to the thin film thickness.

Furthermore, continuous broadening of columns in the orthogonal direction to the deposition plane, so-called \textit{fanning effect}, is directly linked to the ad-atom diffusion ability of incident material particle flux. Once the incident particle flux has condensed on the sample surface, further mass transport might occur by diffusive processes which can be a significant factor in thin film growth, and might potentially alter the shape, size, and orientation of growing columns. The material dependent continuous evolution of porosity and therefore film density affect the values of anisotropic depolarization parameters, accordingly. 
However, the diffusion of incident particle flux can be different along $\mathrm{a}$-axis from that of $\mathrm{b}$-axis, those axes ($\mathrm{a}$,$\mathrm{b}$) form the deposition plane. Such anisotropic diffusion mechanism results in a difference between the radius along $\mathrm{a}$ and $\mathrm{b}$ axes (see Figure~\ref{Fig2_schematics_AB-EMA}~(c) depicted as gray hatched plane).\cite{macnally2020glancing,maudet2020optical}
This sometimes results in the coalescing of individual columns and fibroids which may lead to the formation of bulk-thin-film.\cite{maudet2020optical} The top view HR-SEM images of fabricated SCTFs shown in Figure~\ref{Fig2_Growth_SEMs} verify the existence of column extinction and coalescence. In the supplementary material, we also provide the behavior of retrieved $q_{\mathrm{a}}$ parameter with respect to the thickness of nanocolumnar structure. Except from NiFe SCTFs, for the other material types, the increase in the thin film thickness leads to a decrease in $q_{\mathrm{a}}$. For NiFe SCTFs case, similar to the behavior of parameter $q_{\mathrm{b}}$, the pronounced fanning of column diameter leads to the parameter $q_{\mathrm{a}}$ asymptotically reaches out the saturation state faster as compared to the other material choices (see Figure~S3 in the supplementary material).  As a summarizing figure of this comprehensive work, Figure~\ref{fig_qc_eff}, shows the evolution of effective depolarization value along $\mathrm{c}$ axis versus normalized thickness. Within this representation, the depolarization factor along the major polarizability axis, $\mathrm{c}$, exhibits a material independent, i.e., universal, inverse column length dependence.

\section{Conclusion}

The present work elucidates the inherent optical properties of complex sculptured thin films. The structural parameters of SCTF structures such as film thickness and slanting angle obtained from Mueller matrix SE data analysis show good agreement with those obtained from HR-SEM image analysis. To gain an in-depth understanding on the behavior of depolarization parameters, we fabricated SCTFs from a wide variety of material types. More importantly, we investigated the correlation in between the depolarization factors and the structural parameters of SCTFs. The proposed set of semi-empirical analytical relations for depolarization factors can extrapolate the findings for thicker/thinner films scenarios. 
As another segment of this comprehensive study, we obtained the thickness dependent behavior of birefringence and dichroism properties. We successfully monitored one-dimensional characteristic of the fabricated nanocolumnar structures from a ultra-wide-bandgap (ZrO$_{2}$), a low band-gap (Si), a zero-bandgap (Ti), and a Drude magnetic metal alloy (NiFe) materials over a broad spectral range from near-IR to vacuum-UV. We therefore investigated the presence of a material dependent universal characteristics of the depolarization factors. 
The systematic AB-EMA model based Mueller matrix SE data analysis demonstrated the presence of a generalized thin film thickness dependent behavior of depolarization factor along all polarizability axes. We observed that the depolarization factors are extremely sensitive to the changes in critical dimensions (e.g., radius and thickness of the columnar structure) of the nanostructure platforms. Hence, we predict that these parameters can be utilized in the accurate evaluation and precise optimization of optical responses 
of next-generation sensors with high penetrability for liquid and gaseous substances, waveguide systems, and also photonic integrated circuit applications, for example.

\section{Experimental Section}
\subsection{Fabrication of SCTFs}

In order to deposit SCTFs, we used a custom built, ultra-high vacuum GLAD system with a base pressure of 1.0$\times$10$^{-8}$ mbar. The nanostructures were grown on low p-type doped (100) oriented single crystal Si substrates. The substrates have a native oxide layer of approximately 1.8~nm measured prior to the deposition via generalized SE technique. An electron beam evaporation was utilized and the particle flux impinges on the substrate surface with an oblique angle ($\theta_{\mathrm{flux}}$) of 85$^{\circ}$ from the surface normal. A schematic illustration of glancing angle deposited columnar structure procedure is displayed in Figure~\ref{Fig1_Growth}. The integration of quartz crystal microbalance (QCM) deposition controller to the deposition chamber enables the real-time monitoring of both the deposition rate (in \AA/s) and final thickness of deposited material (in \AA). Table~\ref{Table:GLADparameters} lists all other important parameters that were used in deposition of these materials.

It is important to note that during the deposition of each SCTF, we aimed at maintaining the deposition rate constant by controlling the electron beam current value. Table \ref{Table:GLADparameters} shows the average electron beam parameters and resulting deposition rate values for each material choice. In coordination with QCM device, the electronic shutter system of deposition chamber allows for the precise control on when to start and end the deposition. 
By taking the advantage of high-resolution scanning electron microscopy (HR-SEM) technique, the cross section and top views of columnar structures were obtained (see Figure~\ref{Fig2_Growth_SEMs}). The structural parameters (thickness, diameter, and slanting angle) were also determined from HR-SEM image analysis by using commercially available software package \textit{ImageJ} program.\cite{schneider2012nih} As a complimentary study, we also performed X-ray diffraction (XRD) scans of our fabricated SCTFs (see Figure~\ref{FigX_XRD}). XRD measurements were performed with a Rigaku Smart Lab Diffractometer using $Cu-K\alpha$ radiation. We found out that while NiFe, Ti, and ZrO$_{2}$ SCTFs have poly-crystalline nature, 
Si SCTFs are in amorphous state. Based on the further analysis of XRD data, we found out that ZrO$_{2}$ SCTFs are in cubic phase.\cite{waghmare2018spray} Even if the observed relative intensity XRD peaks are weaker as they are compared with ZrO$_{2}$ case, NiFe and Ti SCTFs were found in face centered cubic.\cite{gupta2008magnetization} and hexagonal closed packed \cite{arshi2013thickness} phases, respectively.

\subsection{Optical Characterization}

The \textit{ex-situ} spectroscopic ellipsometry data acquisitions in Mueller matrix configuration within the spectral range from 0.72~eV to 6.5~eV using a dual rotating compensator ellipsometer (RC2, J.A. Woollam Co., Inc.) were performed for each thin film. The spectrum of Mueller matrix SE data were collected at angles of incidence ($\phi_{\mathrm{AOI}}$) of 45$^{\circ}$, 55$^{\circ}$, 65$^{\circ}$, and 75$^{\circ}$. At each angle of incidence, the spectrum was also measured over a full azimuthal rotation ($\phi_{\mathrm{Azimuth}}$) from 0$^{\circ}$ to 360$^{\circ}$ with increments of 6$^{\circ}$. The resulting data set were then analyzed using the ellipsometry model software (\textit{WVASE} by J.A. Woollam Co., Inc.). As an example, the measured and the best-match model obtained Mueller matrix elements were plotted with respect to both azimuthal orientation of columns and photon energy spectrum (see the supplementary material Figure S1 and S2, respectively).

\subsubsection{\label{sec:MM-GSE}  Mueller Matrix Spectroscopic Ellipsometry}
Spectroscopic ellipsometry is a widely-used non-contact, non-destructive, and non-invasive optical metrology technique. Due to its aforementioned capabilities, this indirect, model driven technique became superior in determining various optical and structural properties of highly anisotropic nanostructured metamaterial platforms.\cite{schmidt2013anisotropic,gospodyn2006characterization,kilic2021broadband}

Within the Stokes descriptive system, the Mueller matrix elements (M$_{ij}$ (where i and j: 1, 2, 3, and 4)) relate the Stokes vector of incoming beam (S$_{\mathrm{in}}$) to that of outgoing beam (S$_{\mathrm{out}}$) after its interaction with the sample under investigation. This relationship between S$_{\mathrm{in}}$ and S$_{\mathrm{out}}$ is given as follows:\cite{tompkins2005handbook,Fujiwara_2007}   

\begin{equation}
\label{eq:muellermatrix1}
\left( {{\begin{array}{*{20}c}
 {S_{0} } \hfill \\ {S_{1} } \hfill \\  {S_{2} } \hfill \\  {S_{3} } \hfill \\
\end{array} }} \right)_{\mathrm{out}} =
\left( { { \begin{matrix} { M_{ 11 } } & M_{ 12 } & M_{ 13 } & M_{ 14 } \\ M_{ 21 } & M_{ 22 } & M_{ 23 } & M_{ 24 } \\ M_{ 31 } & M_{ 32 } & M_{ 33 } & M_{ 34 } \\ M_{ 41 } & M_{ 42 } & M_{ 43 } & M_{ 44 } \end{matrix} } } \right) 
\left( {{\begin{array}{*{20}c}
 {S_{0} } \hfill \\ {S_{1} } \hfill \\  {S_{2} } \hfill \\  {S_{3} } \hfill \\
\end{array} }} \right)_{\mathrm{in}}.
\end{equation}
Each Stokes vector component ($S_{0}$,~$S_{1}$,~$S_{2}$,~$S_{3}$), which is defined in Equation \ref{eq:muellermatrix1}, has the following relations; $S_{0}=I_{\mathrm{p}}+I_{\mathrm{s}}=I_{45}+I_{-45}=I_{+}+I_{-}$, $S_{1}=I_{\mathrm{p}} - I_{\mathrm{s}}$, $S_{2}=I_{45}-I_{ -45}$, and $S_{3}=I_{ + }-I_{ - }$. Here, $I_{\mathrm{p}}$, $I_{\mathrm{s}}$, $ I_{45}$, $I_{-45}$, $I_{ + }$, and $I_{-}$ denote the intensities for the $\mathrm{p}$-, $\mathrm{s}$-, +45$^{\circ}$, -45$^{\circ}$, left handed, and right handed circularly polarized light components, respectively.

\subsubsection{\label{sec:AB-EMA} Anisotropic Bruggeman Effective Medium Formalism}

\begin{figure}
\centering
\includegraphics[width=0.65\linewidth]{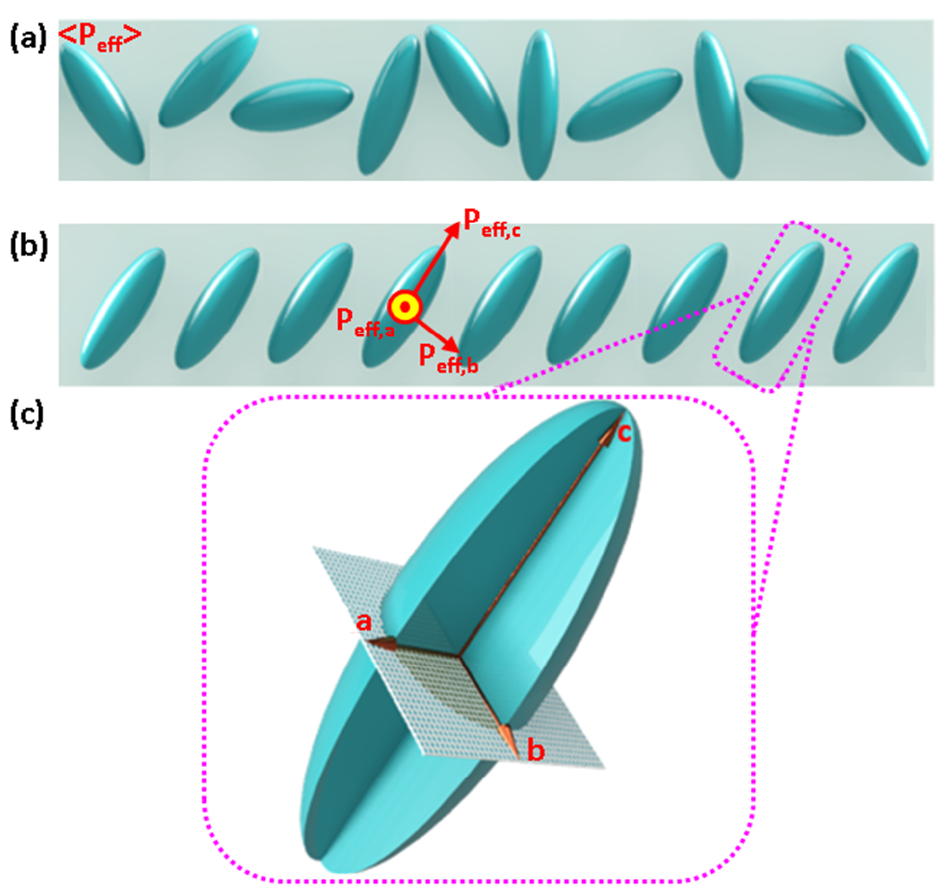}%
\caption{Two different effective medium scenarios which are made up of ellipsoidal inclusions (i.e., the prolate spheroid) in a homogeneous host medium. (a) The ellipsoid inclusions are randomly oriented which exhibits an average effective polarizability $P_{\mathrm{eff}}$. (b) The highly spatially coherent ellipsoid inclusions lead to the emergence of anisotropic properties and therefore the definition of three effective polarizabilities $P_{\mathrm{eff},j}$ where \textit{j}:$\mathrm{a}$,$\mathrm{b}$, and $\mathrm{c}$. (c) The major polarizability axes ($\mathrm{a}$, $\mathrm{b}$, and $\mathrm{c}$) which renders the biaxial nature of the film, has been displayed over a single ellipsoid inclusion. }
\label{Fig2_schematics_AB-EMA}
\end{figure}

\begin{figure*}
\centering
\includegraphics[width=1\textwidth]{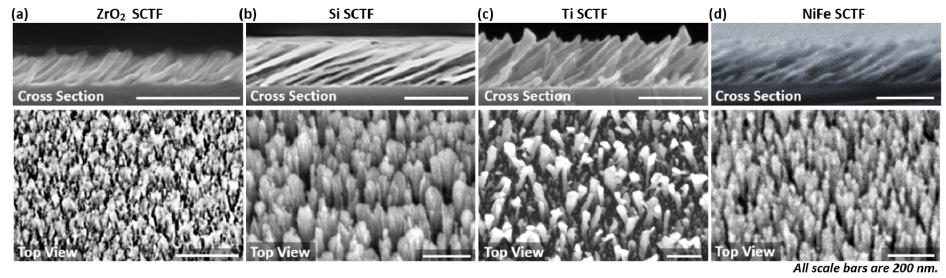}%
    \caption{ Cross-sectional and top views of high-resolution SEM images of the slanted columnar thin films from (a) zirconia (ZrO$_{2}$), (b) silicon (Si), (c) titanium (Ti), and (d) permalloy (NiFe) on crystalline silicon substrates. Scale bars are 200 nm.}
\label{Fig2_Growth_SEMs}
\end{figure*}

\begin{figure}
\centering
    \includegraphics[width=0.75\linewidth]{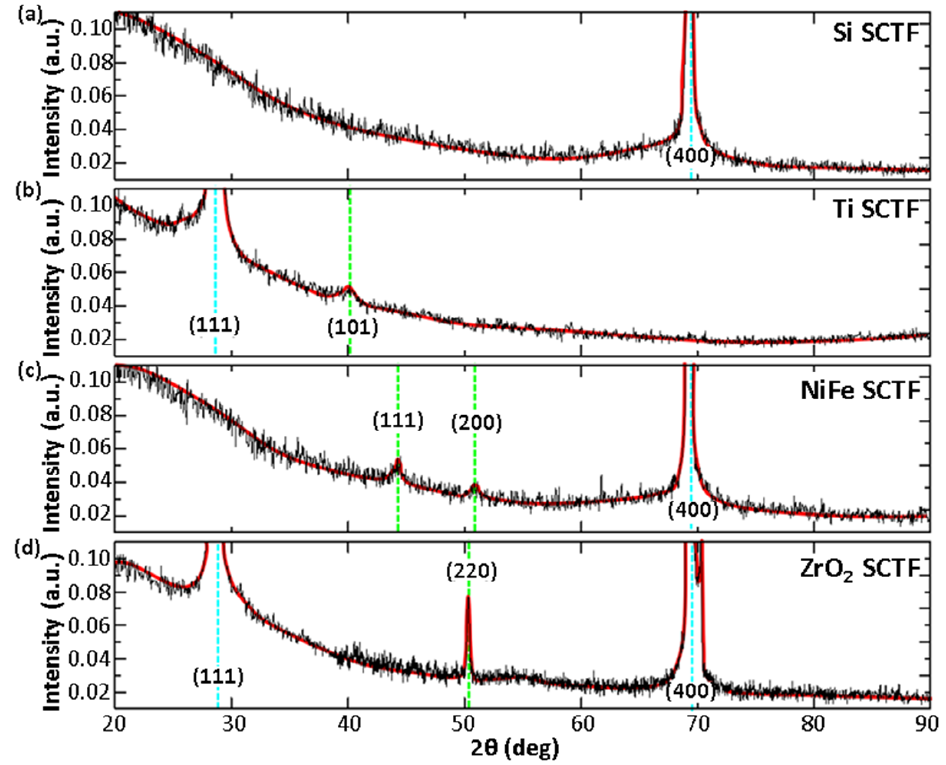}%
    \caption{X-ray diffraction scans of (a) Si ($d=148$~nm), (b) Ti ($d=151$~nm), (c) NiFe ($d=128$~nm), and (d) ZrO$_{2}$ ($d=238$~nm) SCTFs. XRD scans were performed at a fixed angle equal to the slanting angle ($\theta_{s}$) of columnar structures that is obtained from AB-EMA based MM-SE data analysis which are obtained as 49$^{\circ}$, 36$^{\circ}$, 28$^{\circ}$, and 51$^{\circ}$ for Si, Ti, NiFe, and ZrO$_{2}$ SCTFs, respectively. The red solid lines behind each XRD spectrum is placed for guiding the eye. }
\label{FigX_XRD}
\end{figure}

\begin{figure*}[btp]
\centering
\centering
    \includegraphics[width=0.95\textwidth]{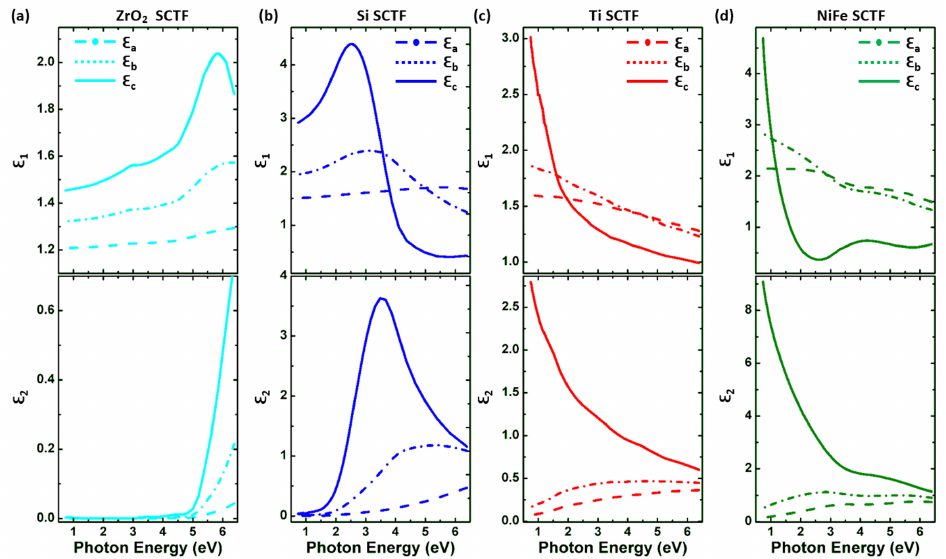}%
    \caption{Complex dielectric function ($\varepsilon=\varepsilon_{1}+i\varepsilon_{2}$)  spectra along each major polarizability axis ($\mathrm{a}$,$\mathrm{b}$,$\mathrm{c}$), for (a) ZrO$_{2}$, (b) Si, (c) Ti, and (d) NiFe SCTFs which are determined from AB-EMA based Mueller matrix SE data analysis within the spectral range from 0.72~eV to 6.5~eV.   
}
\label{Fig_dielectric}
\end{figure*}

\begin{figure*}
\centering
\centering
    \includegraphics[width=0.95\textwidth]{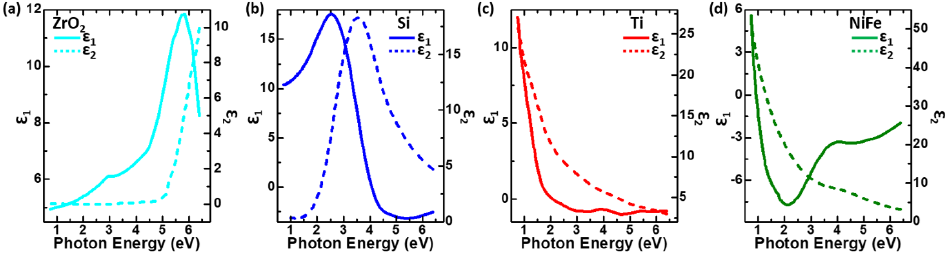}%
    \caption{Constituent bulk-like optical constants, real ($\varepsilon_{1}$) and imaginary ($\varepsilon_{2}$) parts of dielectric functions, of  (a) ZrO$_{2}$,  (b) Si, (c) Ti, and  (d) NiFe SCTFs as determined by traditional AB-EMA within the spectral range from 0.72~to 6.5~eV.}
\label{Fig_bulkdielectric}
\end{figure*}

\begin{figure*}
\centering
\centering
\includegraphics[width=1\textwidth]{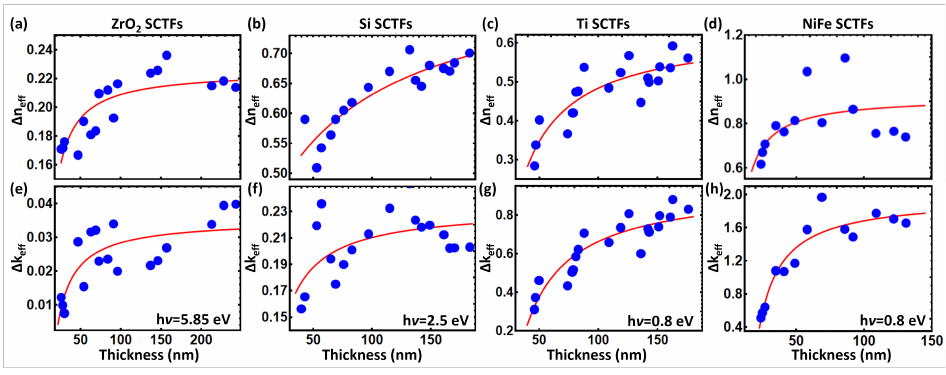}%
\caption{The effective birefringence ($\Delta~n_{eff}=(n_{\mathrm{c}}-(n_{\mathrm{b}}+n_{\mathrm{a}})/2)$) and dichroism ($\Delta~k_{eff}=(k_{\mathrm{c}}-(k_{\mathrm{b}}+k_{\mathrm{a}})/2)$) values (blue circle symbols) at selected photon energy where the refractive index along major polarizability axes (n$_{c}$) is maximum. The resulting $\Delta~n_{eff}$ and $\Delta~k_{eff}$ are shown in (a and e) for ZrO$_{2}$, (b and f) for Si, (c and g) for Ti, and (d and h) for NiFe SCTFs, respectively.   
The red solid lines are obtained based on the asymptotic semi-empirical relation (see Equation \ref{eq_BireDich}).  
}
\label{Fig_Dneff_Dkeff}
\end{figure*}

\begin{figure*}[!htb]
\centering
\centering
\includegraphics[width=0.9\textwidth]{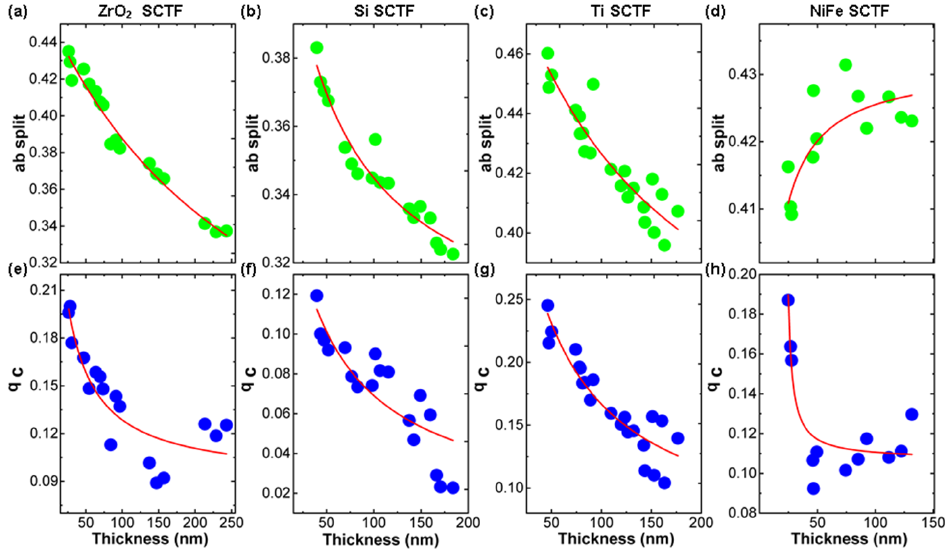}%
\caption{Unraveling aspect ratio driven universal correlation mechanism of anisotropic depolarization factors.
The split between depolarization factors along $\mathrm{a}$-axis (q$_{\mathrm{a}}$) and $\mathrm{b}$-axis (q$_{\mathrm{b}}$) so-called q$_{\mathrm{a}\textit{b}}$ and the depolarization factor along $\mathrm{c}$-axis (q$_{\mathrm{c}}$) are presented. These two were the fitting parameters of the AB-EMA based Mueller matrix  SE data analysis model and were applied to multiple SCTF samples of varying thicknesses for (a and e)  ZrO$_{2}$, (b and f) Si, (c and g) Ti, and (d and h) NiFe. The red solid lines are the fitting curves for each data set. 
}
\label{Fig_depolarization}
\end{figure*}

\begin{figure*}[!ht]
\centering
\includegraphics[width=1\textwidth]{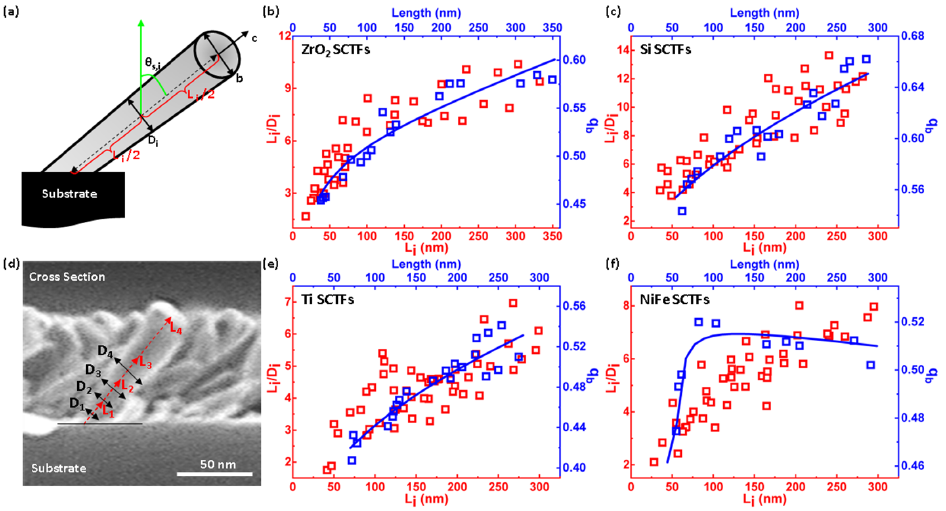}
\caption{
(a) Schematic illustration of an isolated slanted columnar structure with the labels of length and the corresponding diameter values (L$_{i}$,D$_{i}$) along b-axis. Evolution of the aspect ratio given by the column length (L$_{i}$) divided by the column diameter (D$_{i}$) along b-axis obtained from HR-SEM image analysis (red square symbols) as a function of column length that is plotted with the average length of columns  evolution of $q_{\mathrm{b}}$ (blue square symbols) for (b) ZrO$_{2}$, (c) Si, (e) Ti, and (f) NiFe SCTFs. Blue solid line is the extracted fitting curve that conforms Equation \ref{Eq:qa-qb}, which is derived from the unity equation of depolarization factors (Equation \ref{Eq:sum_frac}). (d) cross section HR-SEM image of an example isolated nanocolumnar structure. 
}
\label{figure_qb}
\end{figure*}

\begin{figure}[htb]
\centering
\includegraphics[width=0.65\linewidth]{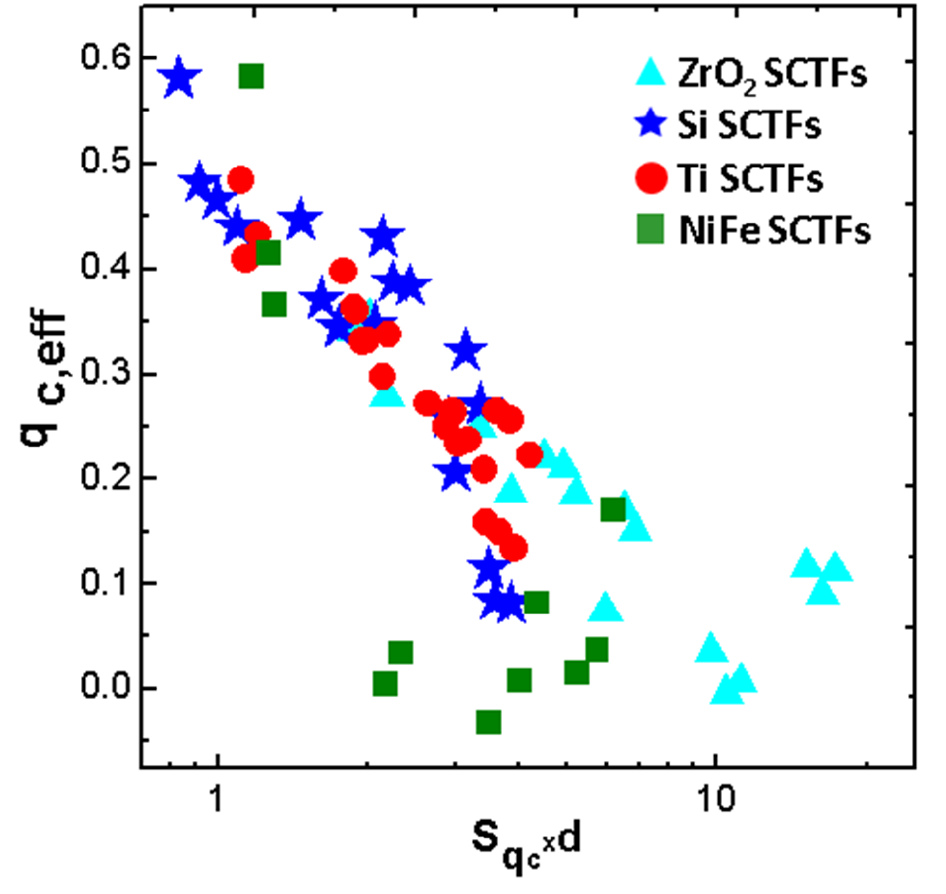}
\caption{
Universal inverse evolution of effective depolarization factor along c-axis (q$_{c,eff}=(q_{\mathrm{c}}-A_{q_{\mathrm{c}}})/q_{0}$) as a function of the normalized column length ($S_{q_\mathrm{c}} ~d$) for different material choice (ZrO$_{2}$(cyan triangle symbol), Si (blue star symbol), Ti (red circle symbol), and NiFe (green square symbol) SCTFs ).  
}
\label{fig_qc_eff}
\end{figure}

The Bruggeman effective medium approximation is a homogenization approach to describe an effective dielectric response of structured materials with inclusions in a host matrix. As an example, a random alignment of ellipsoids in a medium is depicted in Figure~\ref{Fig2_schematics_AB-EMA}~(a). In this case, the effective medium possesses an isotropic effective polarizability $<P_{\mathrm{eff}}>$ and dielectric function $\varepsilon _{\mathrm{eff}}$. 

For spatially aligned and elliptically elongated inclusions, AB-EMA permits description of an effective anisotropic response within an orthorhombic basis system \cite{schmidt2013anisotropic}. Figure \ref{Fig2_schematics_AB-EMA} (b) shows an illustration of the ordered arrangement of ellipsoidal inclusions within a host medium with three effective polarizability directions ($P_{\mathrm{eff},\mathrm{a}}$,~$P_{\mathrm{eff},\mathrm{b}}$,~$P_{\mathrm{eff},\mathrm{c}}$). AB-EMA is a valid and demonstrated approach to render the optical properties of SCTFs. Thereby, three effective major dielectric functions ($\varepsilon_{\mathrm{eff},j}$ where $j=\mathrm{a},\mathrm{b},\mathrm{c}$) are introduced for a mixture of $m$ constituents

\begin{equation}
\sum _{ n=1 }^{ m }{ { f }_{ n }\left( \frac { { \varepsilon  }_{ n }-{ \varepsilon  }_{ \mathrm{eff},j } }{ { \varepsilon  }_{ \mathrm{eff},j }+{ q }_{ j }\left( { \varepsilon  }_{ n }-{ \varepsilon  }_{ \mathrm{eff},j } \right)  }  \right)  } =0 ,
\label{Eq:ab_EMA}
\end{equation}

\noindent where $\varepsilon _{n}$ is the dielectric function of the respective bulk material, $q_{ j }$:~($q_{ \mathrm{a} }$,~$q_{ \mathrm{b} }$,~$q_{ \mathrm{c} }$) are the depolarization factors along the three orthogonal major polarizability axes ($\mathbf{a}$, $\mathbf{b}$, and $\mathbf{c}$), and which are given as follows;\cite{schmidt2013anisotropic}

\begin{equation}
q_{j}=\frac{U_{\mathrm{a}}U_{\mathrm{b}}U_{\mathrm{c}}}{2}\int_{0}^{\infty}\frac{ s+U_{j}^{2}}{\sqrt{\prod_{i}^{\mathrm{a,b,c}}(s+U_{i}^{2})}}ds.
\label{Eq:depol_fac}
\end{equation}

The real-valued depolarization factors (\textit{q}$_{j}$) depend on the real-valued shape parameters ($U_{j}$) of the ellipsoid. The sum of all fraction parameters ($f_{n}$) and the sum of all depolarization factors must equal unity

\begin{equation}
\sum _{ n=1 }^{ m }{ f_{ n } }=f_{ 1 }+f_{ 2 }+f_{ 3 }...+f_{ n } =1,
\label{Eq:sum_frac1}
\end{equation}
\begin{equation}
q_{\mathrm{a}}+q_{\mathrm{b}}+q_{\mathrm{c}}=1.
\label{Eq:sum_frac}
\end{equation}

The diagonalized dielectric tensor ($\varepsilon$) along the major polarizability axes ($\mathrm{a}$, $\mathrm{b}$, and $\mathrm{c}$),
takes the following form
\begin{equation}
\varepsilon=\begin{pmatrix}
\varepsilon_{\mathrm{a}} & 0 & 0 \\
0 &\varepsilon_{\mathrm{b}}  & 0 \\
0 & 0 & \varepsilon_{\mathrm{c}}
\end{pmatrix}.
\label{abEMA}
\end{equation}

We further introduce the ``$\mathrm{a}-\mathrm{b}$" 
split parameter, $q_{\mathrm{ab}}$, that is given as follows;

\begin{equation}
q_{\mathrm{ab}}=\frac{q_\mathrm{a}}{q_\mathrm{a}+q_\mathrm{b}}.
\label{eq:qab}
\end{equation}

We then use $q_{\mathrm{c}}$ and $q_{\mathrm{ab}}$ as variable parameters in the AB-EMA model. 
Furthermore, 
the corresponding analytical relations for both q$_{\mathrm{a}}$ and q$_{\mathrm{b}}$ can trivially be retrieved from both q$_{\mathrm{ab}}$ and q$_{\mathrm{c}}$ relations and are given as follows; 
\begin{equation}
q_{\mathrm{b}}=(1-q_{\mathrm{ab}})(1-q_{\mathrm{c}}) \quad q_{\mathrm{a}}=q_{\mathrm{b}}\left( \frac{q_{\mathrm{ab}}}{1-q_{\mathrm{ab}}} \right) 
\label{Eq:qa-qb}
\end{equation}
However, parameter $q_{\mathrm{ab}}$ is a convenient measure to judge the behavior of elongated structures such as SCTFs. Possible scenarios for q$_{\mathrm{ab}}$ and its effect on individual depolarization factors (q$_{\mathrm{a}}$, q$_{\mathrm{b}}$, q$_{\mathrm{c}}$) are listed below.
\begin{itemize}

\item{$q_{\mathrm{ab}}\rightarrow 1$} Conditions arise such that $q_{\mathrm{a}}\rightarrow 1$, $q_\mathrm{b}\rightarrow 0$, $q_\mathrm{c} \rightarrow 0$, or $q_\mathrm{b} \rightarrow 0$, $0<q_\mathrm{a}<1$, $0<q_\mathrm{c}<1$. The former case represents infinitely thin sheets extended within the $(\mathrm{b}-\mathrm{c})$ plane, the latter represents the case of infinitely thin rods oriented along direction $\mathrm{b}$ with elliptical cross section. \\

\item{$q_{\mathrm{ab}}\rightarrow 0$}  In this case we must require that $q_\mathrm{b} > 0$. Then $q_\mathrm{c}$ may take any value between zero and $1-q_\mathrm{b}$. Then, we can conclude that $q_\mathrm{a}\rightarrow 0$. Hence, the case describes when either infinitely long rods with elliptical ($q_\mathrm{b} \ne q_\mathrm{c}$) or circular ($q_\mathrm{b} = q_\mathrm{c}$)  cross section form along direction $\mathrm{a}$ where $q_\mathrm{b} > 0$ and $q_\mathrm{c}>0$, or when infinitely thin sheets form in either plane $\mathrm{a-b}$ ($q_\mathrm{b} \rightarrow 0$, $q_{\mathrm{c}}\rightarrow 1$) or $\mathrm{a-c}$ ($q_\mathrm{b} \rightarrow 1$, $q_c\rightarrow 0$). \\

\item{$q_{\mathrm{ab}}\rightarrow\frac{1}{2}$} In this case $1>q_{\mathrm{a}}\approx q_{\mathrm{b}}>0$. Hence, $q_{\mathrm{c}}\approx 1-2q_{\mathrm{a}}$. When $q_{\mathrm{c}}<(<<) q_{\mathrm{a}}$ then a (an infinitely) long cylinder with circular cross section is described with direction along $\mathrm{c}$. When $q_{\mathrm{c}}>(>>) q_{\mathrm{a}}$ then a (an infinite) sheet is described within plane $\mathrm{(a-b)}$.\\

\item{$q_{\mathrm{ab}}\lessapprox\frac{1}{2}$} Then $q_{\mathrm{b}}>\frac{1}{2}(1-q_{\mathrm{c}}$). If we assume that the long axis of the inclusion is along direction $\mathrm{c}$ then its elliptical cross section is described with larger elliptical extend along direction $\mathrm{b}$ than $\mathrm{a}$.\\

\item{$q_{\mathrm{ab}}\gtrapprox\frac{1}{2}$} Then $q_{\mathrm{b}}<\frac{1}{2}(1-q_{\mathrm{c}}$). If we assume that the long axis of the inclusion is along direction $\mathrm{c}$ then its elliptical cross section is described with larger elliptical extend along direction $\mathrm{a}$ than $\mathrm{b}$.\\

\end{itemize}

As will be discussed below, our structures result in $q_\mathrm{ab}$ within $0.3\dots0.45$ and corresponding values for $q_\mathrm{c}$ approximately half of the $q_\mathrm{ab}$ values. Hence, our columnar structures are generally slightly elliptical in cross section with larger extend within the slanting plane than out of the slanting plane. Note that limiting cases $q_{\mathrm{ab}}\rightarrow 0$ and $q_{\mathrm{ab}}\rightarrow 1$ do not correspond to SCTFs in our case because our nanostructures do not resemble sheets, and we have by definition placed direction $\mathrm{c}$ along the columnar axis, i.e., the direction of the longest extend of our nanostructures. Therefore, situation $d.$ described above is the scenario representative of all our samples.

\medskip
\textbf{Supporting Information} \par 
Supporting Information is available from the Wiley Online Library or from the author.

\medskip
\textbf{Acknowledgements} \par 
This work was partially supported by the National Science Foundation under award number DMR 1808715, Air Force Office of Scientific Research under award number FA9550-18-1-0360, Nebraska Materials Research Science and Engineering Center under award number DMR 1420645, Swedish Knut and Alice Wallenbergs Foundation supporting grant titled ’Wide-bandgap semi-conductors for next generation quantum components’, and American Chemical Society/Petrol Research Fund. R. F. acknowledges the German Research Foundation (DFG) award FE 1532/1-1. C. A. acknowledges partial support by the Office of Naval Research Young Investigator Program (ONR YIP) under award number N00014-19-1-2384. M. S. acknowledges the University of Nebraska Foundation and the J. A. Woollam Foundation for financial support.\\

\medskip



\pagebreak
\listoffigures
\pagebreak

\pagebreak

\begin{table}
\centering
\caption{The list of typical GLAD parameter values for the deposition of ZrO$_{2}$, Si, Ti, and NiFe SCTF structures.} 

\resizebox{0.7\textwidth}{!}{
\begin{tabular}{ccccc}
\hline\hline
{Material} &  e-beam Voltage &  e-beam Current& Initial Pressure  & Growth Rate \\ 
{} & (kV)&(mA)& (mBar) & (\AA/s) \\ 
\hline\hline
ZrO$_{2}$ & $8.9$ & $120$ & $5.0\times10^{-6}$    & 0.35 \\ 
Si & $8.8$ & $210$ & $9.0\times10^{-8}$  & 1.34 \\ 
Ti &  $8.9$ & $160$ & $1.5\times10^{-8}$   & 1.9 \\ 
NiFe & $8.9$ & $250$ & $5.0\times10^{-8}$    & 0.63 \\ 
\hline \hline
\end{tabular}
\label{Table:GLADparameters}
}
\end{table} 

\end{document}